\documentclass[iop]{emulateapj}
\usepackage{natbib,mathtools}
\usepackage[normalem]{ulem}
\pdfoutput=1

\shorttitle{Probing isotropy with SNe Ia}
\shortauthors{Javanmardi et al.}

\begin{document}

\title{Probing the isotropy of cosmic acceleration\\ traced by Type Ia supernovae}

\author{B. Javanmardi\altaffilmark{1}}
\affil{Argelander Institut f\"ur Astronomie der Universit\"at Bonn, Auf dem H\"ugel 71, Bonn, D-53121, Germany}
\affil{Max-Planck-Institut f\"ur Radioastronomie, Auf dem H\"ugel 69, Bonn, D-53121, Germany}
\email{behnam@astro.uni-bonn.de}

\author{C. Porciani}
\affil{Argelander Institut f\"ur Astronomie der Universit\"at Bonn, Auf dem H\"ugel 71, Bonn, D-53121, Germany}

\and

\author{P. Kroupa, and J. Pflamm-Altenburg}
\affil{Helmholtz-Institut f\"ur Strahlen und Kernphysik, Nussallee 14-16, Bonn, D-53115, Germany}

\altaffiltext{1}{Member of the International Max Planck Research School (IMPRS) for Astronomy and Astrophysics at the Universities of Bonn and Cologne.}

\label{firstpage}

\begin{abstract} 
We present a method to test the isotropy of the magnitude-redshift relation
of Type Ia Supernovae (SNe Ia) and single out the most discrepant direction (in terms of the signal-to-noise ratio) with respect to the all-sky data.
Our technique accounts for possible directional variations of the corrections
for SNe Ia and yields all-sky maps of the best-fit cosmological parameters with arbitrary angular resolution.
To show its potential, we apply our method to the recent Union2.1 compilation, building maps with three
different angular resolutions.
We use a Monte Carlo method to estimate the statistical significance with which we could reject
the null hypothesis that the magnitude-redshift relation is isotropic based on the properties
of the observed most discrepant directions.
We find that, based on pure signal-to-noise arguments,  the null hypothesis cannot be rejected at any meaningful confidence level. However, if we also consider that the strongest deviations in the Union2.1 sample closely align with the dipole temperature anisotropy of the cosmic microwave background, we find that the null hypothesis should be rejected at the $95-99$ per cent confidence level, slightly depending on the angular resolution of the study.
If this result is not due to a statistical fluke,
it might either indicate that the SN data have not been cleaned from all possible systematics or even point towards new physics.
We finally discuss future perspectives in the field for achieving larger and more uniform data sets that will vastly improve the quality of the results and optimally exploit our method.
\end{abstract}

\keywords{cosmology:dark energy, supernovae:general, methods:data analysis.}

\section{Introduction}
In 1998, the luminosity-redshift relation (Hubble diagram) of a few tens of Type Ia supernovae (SNe) provided the evidence base for
the accelerated expansion of the
universe \citep{riess98,perlmutter99}. Since then, major efforts have been made to increase the sample size, extend it to higher redshift,
and refine the observational and data-reduction techniques. Current datasets already include several hundreds of objects but the quest for
dark energy drives copious activity in this field.

The control of systematic errors is the key to making the study of  SNe Ia a prime cosmological tool.
\citet{suzuki} state that systematic uncertainties already dominate over the statistical ones in the determination of the cosmological
parameters. Given that systematics will become even more important in the future, a careful scrutiny of all the possible sources of
methodological bias is crucial. In this paper, we focus on the spatial isotropy of the Hubble diagram traced by type Ia SNe.
The standard cosmological model is rooted in the assumption that the Universe is homogeneous and isotropic on large scales.
Hence, SNe Ia are expected to (statistically) obey the same dimming relation in all directions.
There are, however, several phenomena that could introduce anisotropies with different characteristic scales and amplitudes
in the observed expansion rate. To name a few:
dust absorption (both in the Milky Way and in the galaxies hosting the SNe), redshift-space distortions due to large-scale motions,
weak gravitational lensing, the presence of large-scale structures and contamination of the SNe Ia samples.
Detecting these effects and correcting for them would ultimately lead to tighter and less biased constraints on the cosmological
parameters.

At the same time, it is healthy to scrutinise the validity of the standard model of cosmology \citep{kroupa12a,kroupa12,kroupa15,koyama} and its fundamental assumptions, namely those of the cosmological principle. Ruling out cosmic isotropy with high statistical
confidence would lead to a major paradigm shift especially if such a conclusion is confirmed by multiple datasets affected by
different systematics.
In this respect,
the analysis of temperature anisotropies in the Cosmic Microwave Background (CMB) has dominated the scene in the last decade.
The WMAP satellite detected a few large-scale ``anomalies'' that somewhat deviate from the expectations of the standard model that best fits
the data on smaller scales \citep{tegmark,eriksen,hansen,copi10b}.
In brief, the quadrupole and octopole terms are surprisingly planar and 
there is a significant alignment between them. Moreover, their normals lie close to the axis of the CMB dipole. This discovery generated a long lasting debate in the literature
concerning whether or not these features are genuine signs of new physics.

Alternatively they could be due to the influence of data processing, to the imperfect removal of foreground contaminants and secondary astrophysical effects \citep{rassat14}, as well as to a statistical fluke \citep{bennett11}.
The Planck satellite recently confirmed the existence of these alignments \citep{planck} suggesting that they are not artifacts of
the data-reduction pipelines. A satisfactory explanation for the origin of these asymmetries is still not available.

The isotropy of the Hubble diagram for SNe Ia has been repeatedly tested.
\citet{kolatt} used 79 SNe from \citet{riess98} and \citet{perlmutter99} to perform localized fits
within an opening angle of $60^\circ$ around random directions. After expanding
the best-fitting cosmological parameters in low multipoles, they found that no dipole anisotropy was statistically
significant.
Subsequent studies mainly adopted two methods: either they compared Hubble diagrams for pairs of hemispheres and looked for the most
discrepant hemispheric cut \citep[Hemispherical Comparison, e.g.][]{schwarz} or fit a dipole angular distribution \citep[Dipole Modulation Fitting, e.g.][]{cooke}. Other authors looked for angular correlations in SN magnitudes \citep{blom08} or analyzed the magnitude-redshift relation in the context of anisotropic cosmological models \citep{koivisto08a,campanelli}. 
Low-redshift samples were used to estimate the direction and
amplitude of the local bulk flow \citep{bonvin06,schwarz,colin,turnbull,rathaus,feindt,kalus,appleby13,appleby14b}.
At the same time, several authors analysed higher-redshift data to look for large-scale anisotropies
\citep{schwarz,gupta10,cooke,antoniou,mariano,cai,li,campanelli,zhao,heneka,wang14,yang,chang,jimenez}, which is also the aim of our work.
Statistically significant deviations have been detected at low redshift \citep{schwarz}, while no high-redshift study could rule out
isotropy at more than 2 Gaussian standard deviations, $\sigma$. 

In this paper, we present a simple but powerful method to test the isotropy of the luminosity-redshift relation for SNe Ia.
Contrary to most previous studies, our analysis neither searches for hemispheric asymmetries and dipolar patterns nor does it use any other
template anisotropic configuration. For each direction on the celestial sphere $\hat{r}\in S^2$, we derive a set of cosmological parameters by considering only the SNe that lie within an angle $\theta$ from $\hat{r}$. We then build maps of these ``local cosmological parameters'' with different values of $\theta$ and identify the directions associated with the most significant anisotropies taking into account that the number of datapoints used in the fit fluctuates from one direction to another.  For completeness, we consider that the correction for the distance modulus of SNe Ia might also depend on $\hat{r}$ due, for instance, to dust extinction. Therefore, our strategy is able to detect anisotropies generated both by physical effects and by systematics.
Even though our method is best suited for the large SN samples with nearly uniform sky distribution that will become available
in the next decade, we provide an example of its potential by applying it to
the Union2.1 SN Ia compilation \citep{suzuki} from the Supernova Cosmology Project (SCP). We limit our study to redshifts $z\geq 0.2$
in order to minimize the influence of local inhomogeneities and bulk flows.

The rest of this paper is organised as follows. Section 2 describes the main properties of the Union2.1 sample. Our method of analysis
is introduced in Section 3. Results are presented and critically discussed in Section 4. Finally, we conclude in Section 5.

\section{Data}
The Union2.1 compilation \citep{suzuki} collects data for 580 SNe Ia in the redshift range of $0.015\leq z\leq 1.414$.
It combines entries from 19 datasets uniformly analysed after adopting strict lightcurve quality cuts and the SALT2 lightcurve-fitter
\citep{guy07}.
The Union2.1 catalog has been built for dark-energy science and updates the previously released Union \citep{kowalski} and Union2
\citep{amanullah} compilations.
In particular, it contains 14 new SNe discovered in the HST Cluster Supernova Survey (a survey run by the SCP) that pass the Union2 selection
cuts. Ten of these SNe are at $z>1$ which makes the Union2.1 sample ideal for studying isotropy out to the largest possible distances.

\begin{figure}
\includegraphics[scale=0.26]{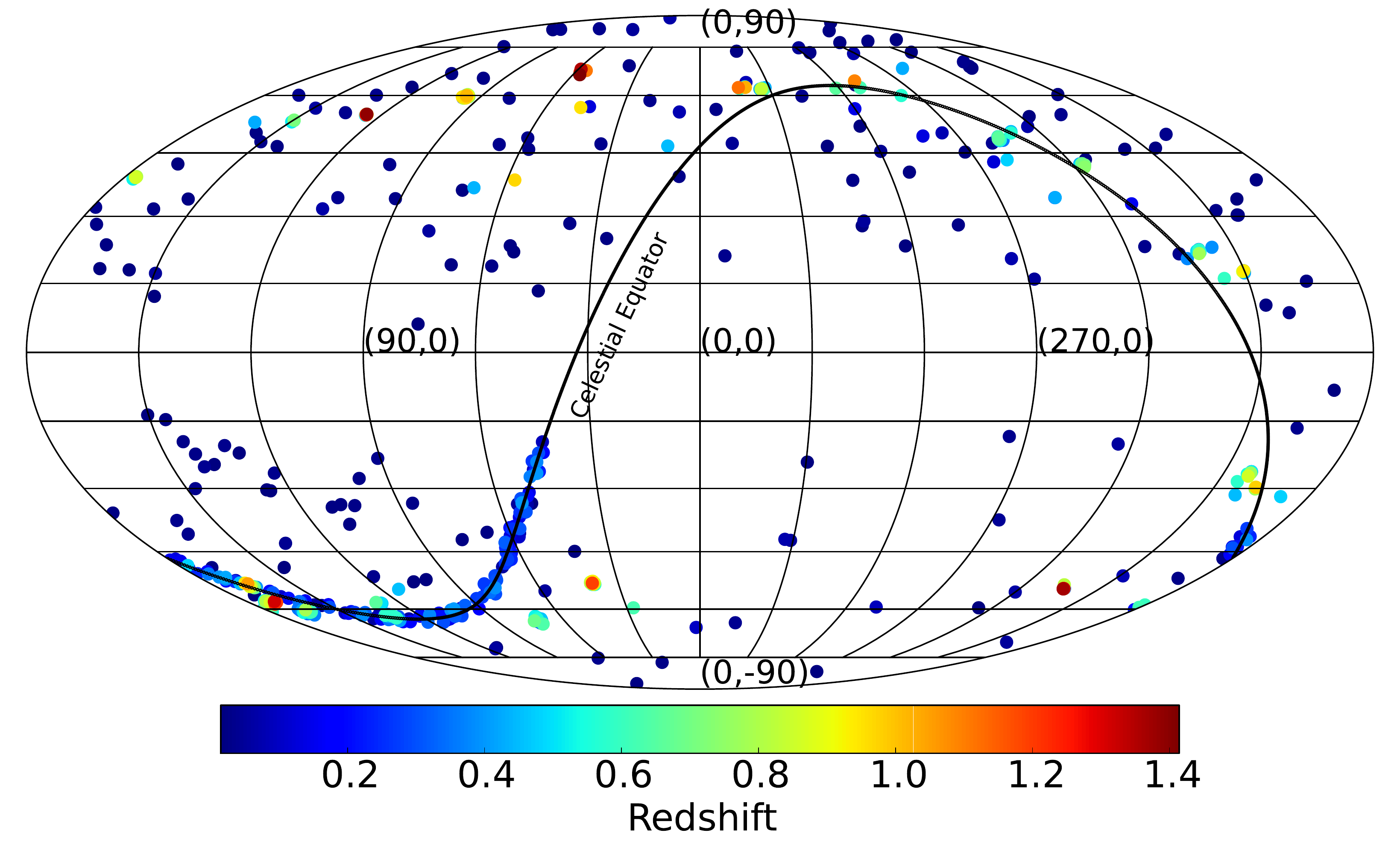}
 \caption{Mollweide projection map of the Union2.1 SNe Ia in the Galactic coordinate system. Each circle corresponds to the position of
a SN Ia on the sky and is colour-coded based on the SN Ia redshift.
The black solid curve indicates the celestial equator.}
\label{fig:molw}
\end{figure}

\begin{figure}
\includegraphics[scale=0.42]{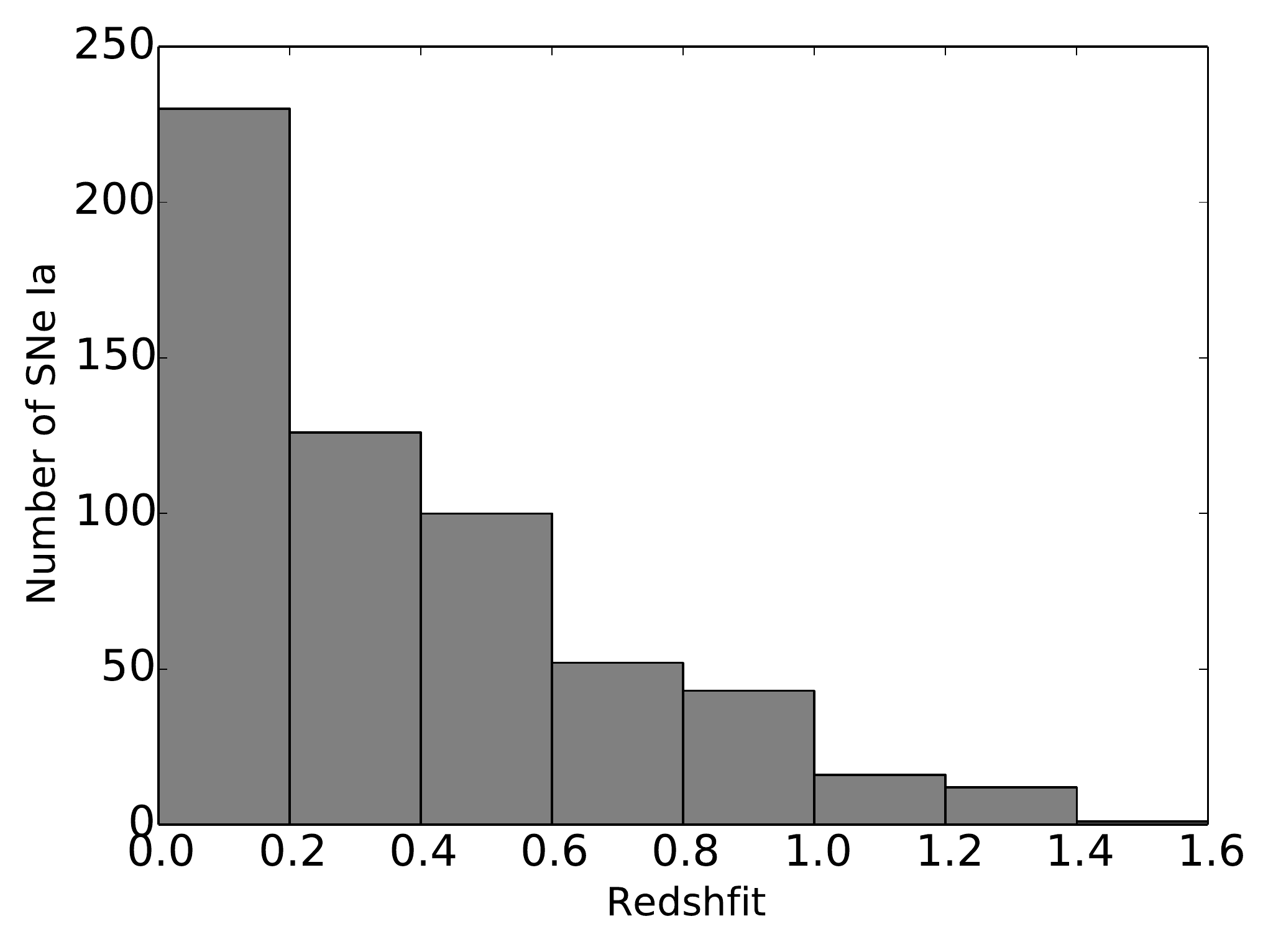}
 \caption{Redshift distribution of Union2.1 SNe.}
 \label{fig:histz}
\end{figure}

The Union2.1 catalog provides five entries for each SN, specifically: name, redshift (CMB centric), distance modulus,
error in the estimate of the distance modulus, and the probability that the SN was hosted by a low-mass galaxy.
We obtained the coordinates for all the SNe Ia in the compilation either from the NASA/IPAC Extragalactic Database
(NED)\footnote{The NASA/IPAC Extragalactic Database (NED) is
operated by the Jet Propulsion Laboratory, California Institute of Technology, under contract with the National Aeronautics and Space
Administration.} or directly from the SuperNova Legacy Survey (SNLS) data release \citep{astier06}.
The sky distribution of Union2.1 SNe is plotted in Figure \ref{fig:molw}. The angular position of each SN is marked by a symbol
which has been colour-coded based on redshift. Several features are immediately apparent in the image. First, there are only a few SNe
close to the galactic plane. Second, an arc-like region in the southern hemisphere is much more densely populated than the rest. This is
the footprint of the Sloan Digital Sky Survey-II (SDSS-II) SN search and corresponds to the southern equatiorial stripe
(Stripe 82, with coordinates
$-50 < RA < 59$ and $-1.25 < DEC < 1.25$) which has been imaged repeatedly with broad wavelength coverage and also been
subject to extensive spectroscopic studies \citep{kessler}.
Finally, high-redshift SNe are very sparsely distributed and rare which is also evident from the redshift distribution of the Union2.1
SNe shown in Figure \ref{fig:histz}.

\section{Method}
As mentioned in the Introduction, previous studies on the isotropy of the luminosity-redshift 
relation of SNe Ia have mainly searched for dipolar anisotropies or 
hemispheric asymmetries. In this section, we introduce a more general method
that does not assume any particular form of the anisotropy.

\subsection{Cone Analysis}\label{sec:cone}
Let us consider a particular direction on the sky, $\hat{r}$, 
with Galactic longitude $l$ and latitude $b$.
In order to single out a finite region surrounding $\hat{r}$, 
we consider a cone with apex angle 
$2\theta$  subtending a solid angle of $\Omega_{cone}=2\pi(1-\cos \theta)$ sr 
on the celestial sphere.
The apex of the cone is located at the centre of the Galactic 
coordinate system and its axis of symmetry points towards $\hat{r}$.
After isolating the SNe Ia contained 
within the cone, we build their magnitude-redshift relation and derive
``local cosmological parameters'' by fitting a theoretical relationship 
to it (see Section \ref{sec:formulation} for details).  
We then vary the cone direction $\hat{r}$ making sure that we cover the whole 
sky. 
For convenience, we move $\hat{r}$ along the pixel centers of a 
HEALPix\footnote{Hierarchical Equal Area isoLatitude Pixelization, http://healpix.sourceforge.net} grid \citep{gorski}.

For the application of the method to the Union2.1 data, we repeat the analysis
using three different opening angles: 
$\theta=\frac{\pi}{2}$ (hemispheres), $\frac{\pi}{3}$ and $\frac{\pi}{6}$.
We use a HEALPix grid with 192 pixels so that the solid angle subtended 
by each pixel is much smaller than that subtended by the cones.

\subsection{Formulation}\label{sec:formulation}
\subsubsection{Global fit}
SNe Ia are not perfect standard candles, their peak brightness correlates with their
color, the light-curve width and the mass of the host galaxy.
In the Union2.1 sample, 
individual lightcurves are analyzed with the SALT2 fitter which provides estimates for three parameters:
the peak magnitude, $m_{obs}$, in the rest-frame B band,
the deviation, $x_1$, from the average light-curve shape and the deviation, $c$, from the mean $B-V$ color.
The color and light-curve-shape corrected distance modulus
is then written in terms of four unknown parameters
($\alpha, \beta, \delta$ and $M_B$) so that
\begin{equation}
\label{eq:muunion}
\mu_{B}=m_{obs}+\alpha\cdot x_{1}-\beta \cdot c+\delta\cdot P_{host}-M_B,
\end{equation} 
where $M_B$ is the absolute B-band magnitude at maximum of a SN Ia and
$P_{host}$ denotes the probability that the SN Ia is hosted by a galaxy with stellar mass
$M_*<10^{10} M_\odot$. This probability is estimated differently for 
untargeted and targeted surveys.

In the context of the theory of general relativity,  homogeneous and isotropic universes are described
by Friedmann-Lema\^{i}tre-Robertson-Walker models. 
For simplicity we only consider flat models 
in which the density parameters for the matter and the cosmological constant satisfy the relation
$\Omega_m+\Omega_\Lambda=1$. 
Following standard practice, we write the magnitude-redshift relation of SNe Ia in terms
of the distance modulus
\begin{equation}
\mu(z)=5\,\log_{10} d_L(z,\Omega_\Lambda)+5\,\log_{10} \left(\frac{D_H}{\mathrm {Mpc}}\right)+25\;,
\end{equation}
where 
\begin{equation}
d_{ L }=\int _{ 0 }^{ z } \frac {(1+z)\, dq }{ \left[{ { (1+q) }^{ 2 }(1+{ \Omega  }_{ m }q)-q(2+q){ \Omega  }_{ \Lambda  } }   \right]^{1/2}}
\end{equation}
is the dimensionless ``Hubble-constant-free" luminosity distance and 
$D_H=c/H_0$ is the Hubble radius defined in terms of the speed of light and the present-day value of
the Hubble constant.

Classically, the SN data are fitted with a cosmological model assuming Gaussian
errors and following a maximum likelihood approach \citep[e.g.][]{astier06}.
For $N$ Type Ia SNe, this corresponds to minimising the target function 
\begin{equation}
\label{eq:chi1}
\chi^2= {\bf V}^T\,{\bf C}^{-1}\,{\bf V} 
\end{equation}
where ${\bf V}$ is a $N$-dimensional vector with elements $V_i=\mu_{B,i}(\alpha,\beta,\delta,M_B)-\mu(z_{i}; H_0,\Omega_\Lambda)$ and ${\bf C}$ is the covariance matrix of the errors in the observed distance moduli. For the Union compilations, this matrix is publicly available. Its off-diagonal elements include several contributions due to the light-curve fits, galactic extinction, gravitational lensing, peculiar velocities and sample-dependent systematics.
The nuisance parameters $\alpha, \beta, \delta$ and $M_B$ are fitted 
simultaneously with the cosmological parameters. Actually, the best-fit values for $M_B$ and $H_{0}$ are completely degenerate as only the combination ${\cal M}=M_B+5 \log_{10}(D_H/{\mathrm {Mpc}})$ appears in eq. (\ref{eq:chi1}). 
Using the whole data set gives the following best-fit values \citep{suzuki} $\alpha=0.121$, $\beta=2.47$, $\delta=-0.032$, and $M_B=-19.321$ (for $H_{0}=70$ km s$^{-1}$Mpc$^{-1}$).

\subsubsection{Local fits}
The parameters $\alpha, \beta$ and $\delta$ describe correlations between different SN observables and might vary for the different surveys of a compilation. \citet{karpenka} found inconsistencies between the values of these correction parameters in the Union2 catalog.
In order to account for possible direction-dependent systematics, when we consider localized sub-samples of the Union2.1 data, we should in principle allow them to vary freely. However, this would require knowledge of the covariance between $m_{obs}$, $x_{1}$ and $c$ for individual SNe. Regrettably this information is not provided in the Union2.1 catalog. We therefore adopt a simplified approach by 
assuming a constant correction, $\mu_{cor}$, for the distant modulus of all the SNe lying within a cone. In other words, we keep the quantities $\alpha, \beta, \delta$ and ${\cal M}$ fixed at their global best-fit value (hereafter denoted with a hat) but we write
\begin{equation}
\label{eq:v1}
V_i=\mu_{B,i}(\hat\alpha,\hat\beta,\hat\delta,\hat{M}_B)-5\log_{10} d_L(z_i;\Omega_\Lambda)-\Delta_{0}
\end{equation}
with $\Delta_{0}=5 \log_{10} (c H_0^{-1})+25-\mu_{cor}$.
Note that $\Delta_{0}$ accounts for both an ``anisotropic Hubble constant" and 
for the mean effect of variations in $\alpha, \beta, \delta$ and $M_B$  due to systematic errors. We are left with a two-dimensional problem. 
For each pixel on the sky, we then determine the best-fitting values of the cosmological parameter
$\Omega_\Lambda$ and of the correction parameter $\Delta_{0}$ by
minimizing the $\chi^2$ target function (covariances are extracted from ${\bf C}$ after identifying the SNe in the cone). However, the model parameters anticorrelate:
directions associated with large values of $\Omega_\Lambda$ provide low values of $\Delta_0$ (and 
viceversa). In order to minimise this effect, we use the luminosity distance, $d_L$, evaluated at the mean redshift of the sample ($\bar{z}=0.36$) as a pivot point and define
\begin{equation}
\label{eq:v2}
V_i=\mu_{B,i}(\hat\alpha,\hat\beta,\hat\delta,\hat{M}_B)-5\log_{10} \left(\frac{d_L(z_i;\Omega_\Lambda)}{d_L(\bar{z};\Omega_\Lambda)}\right)-\Delta
\end{equation}
where $\Delta= \Delta_{0}+5\log_{10} d_L(\bar{z};\Omega_\Lambda)$ and $\Omega_\Lambda$ are our free parameters.

\section{Results}
\subsection{All-sky fit}
To test the consistency of our approach with previous studies, we first perform an all-sky fit.
Results are shown in Table \ref{t:allsky} for the entire Union2.1 sample and for two sub-sets including
the SNe Ia with redshift smaller and larger than $z=0.2$ 
 (in this paper, uncertainties on the value of single parameters always 
 correspond to $\Delta\chi^{2}=1$).
 Our results are in excellent agreement with the analysis in \citet{suzuki} 
 who found  $\Omega_{\Lambda}=0.705^{+0.040}_{-0.043}$ (see their Table 7). Also note that setting
  $\mu_{cor}=0$, $H_{0}=70$ km\,s$^{-1}$Mpc$^{-1}$ and $M_B=-19.321$ corresponds to $\Delta_{0}=43.159$ mag which gives $\Delta=41.419$ mag for $\Omega_{\Lambda}=0.705$ and $\bar{z}=0.36$.

%\clearpage
\begin{table}
\begin{center}
\caption{All-sky fitting results \label{t:allsky}}
\begin{tabular}{cccc}
\tableline
\tableline
&$\Omega_{\Lambda}$&$\Delta$ (mag)& $\chi^2/\nu$\\
 \tableline
All SNe &$0.70_{-0.04}^{+0.04}$ & $41.428^{+0.028}_{-0.031}$&0.94\\

$z \geq0.2$&$0.68_{-0.05}^{+0.06}$ & $41.443^{+0.052}_{-0.049}$&0.94\\

$z < 0.2$&$0.62_{-0.19}^{+0.18}$ & $41.380^{+0.040}_{-0.041}$&0.93\\
 \tableline
\end{tabular}
\tablecomments{Best-fit parameters and the corresponding reduced chi-square, $\chi^2/\nu$, for the entire Union2.1 sample and for two redshift subsets. The quoted uncertainties correspond to $\Delta\chi^{2}=1$.}
\end{center}
\end{table}

\subsection{Cone analysis}
\subsubsection{$\Omega_{\Lambda}$ maps}
\begin{figure*}
\includegraphics[scale=0.4]{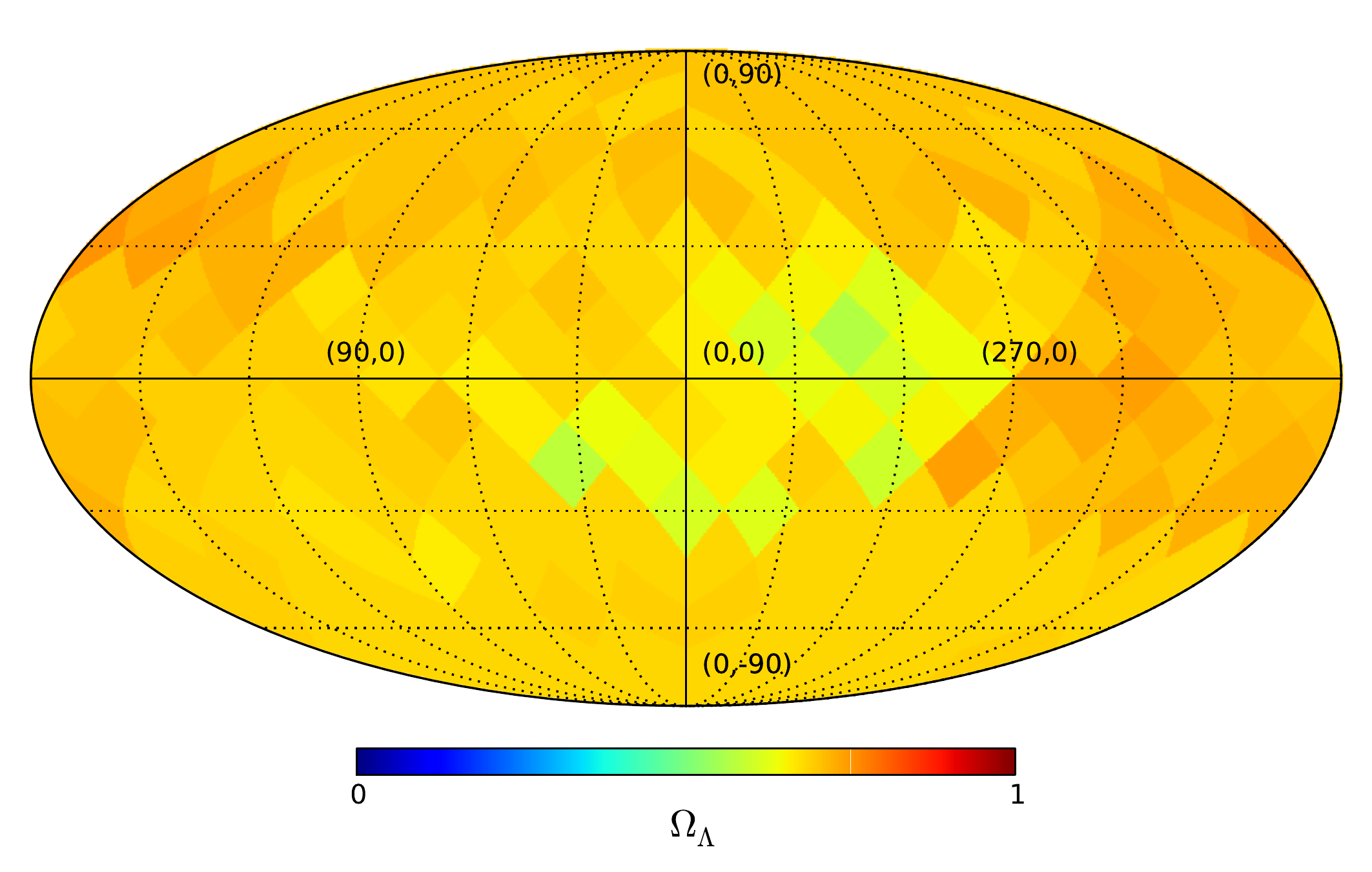}
\includegraphics[scale=0.4]{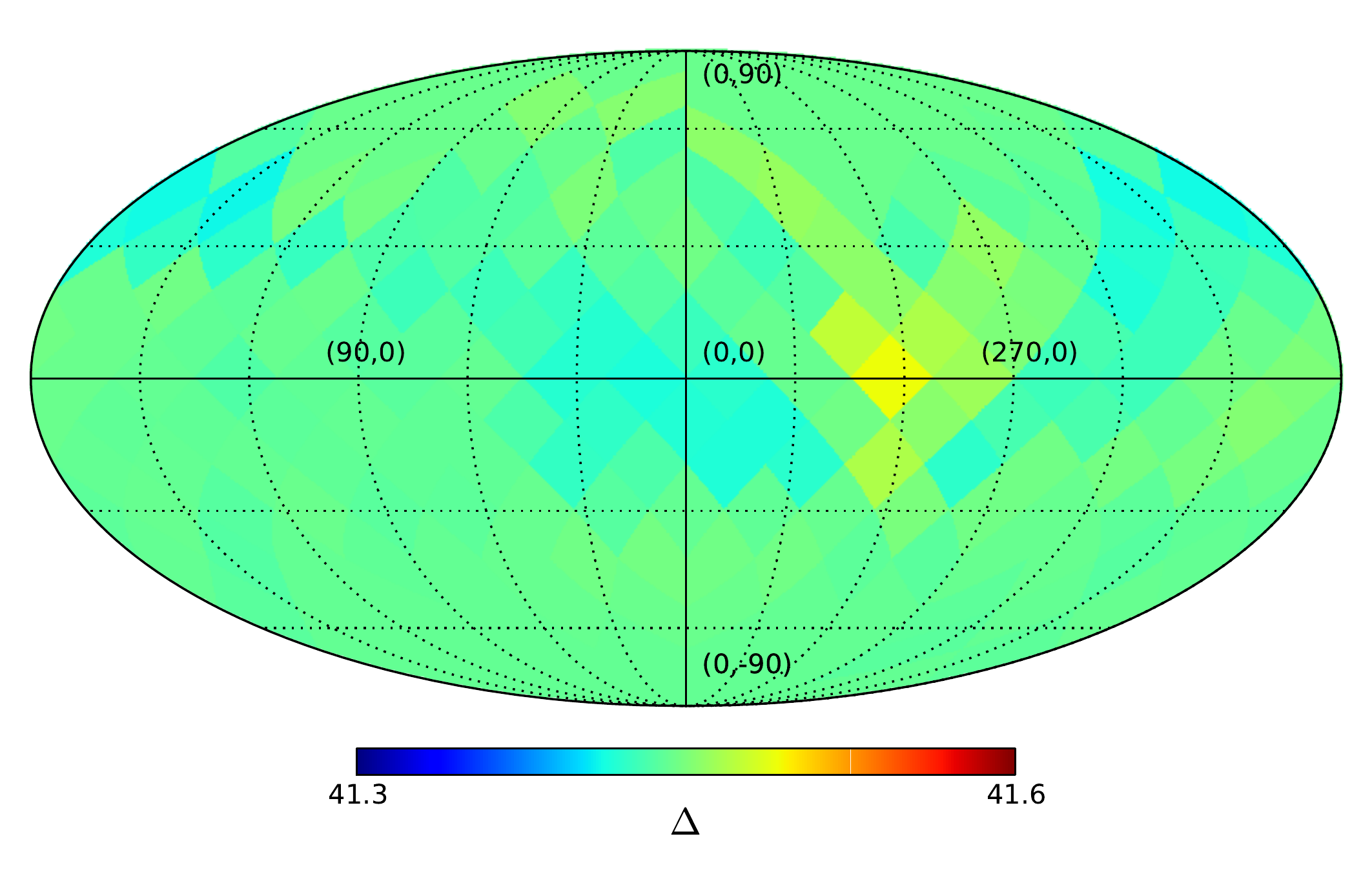}
\includegraphics[scale=0.4]{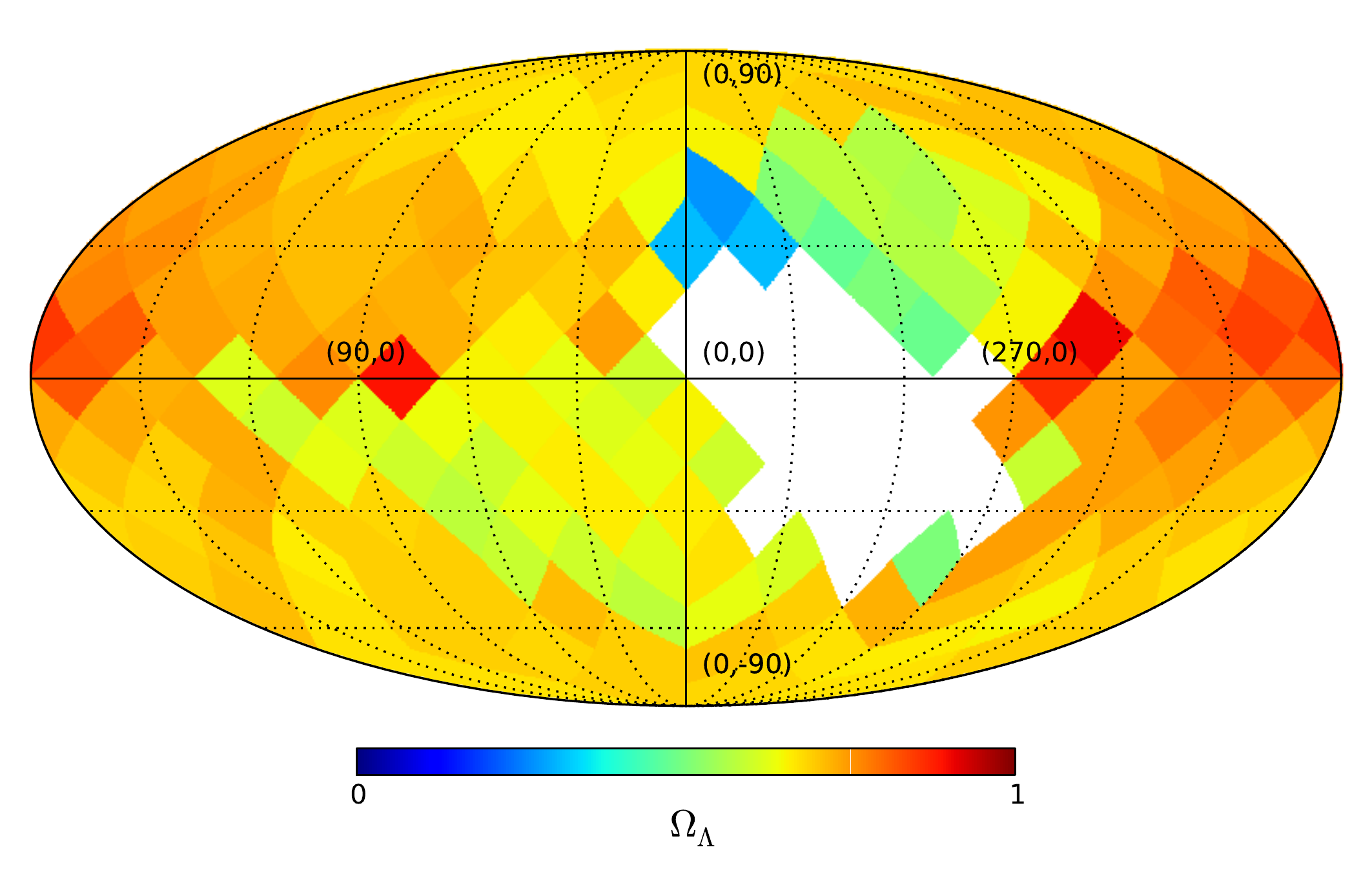}
\includegraphics[scale=0.4]{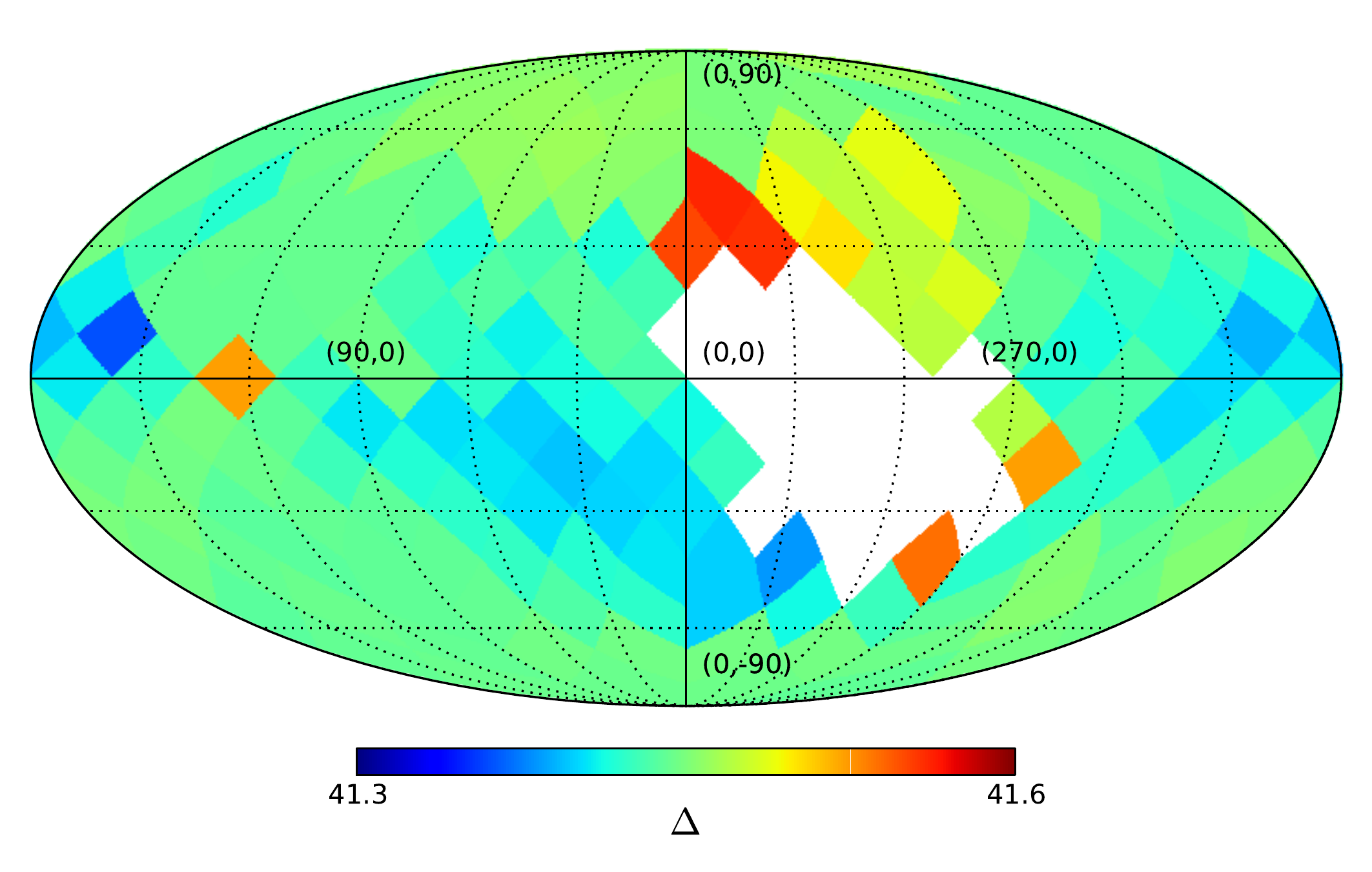}
\includegraphics[scale=0.4]{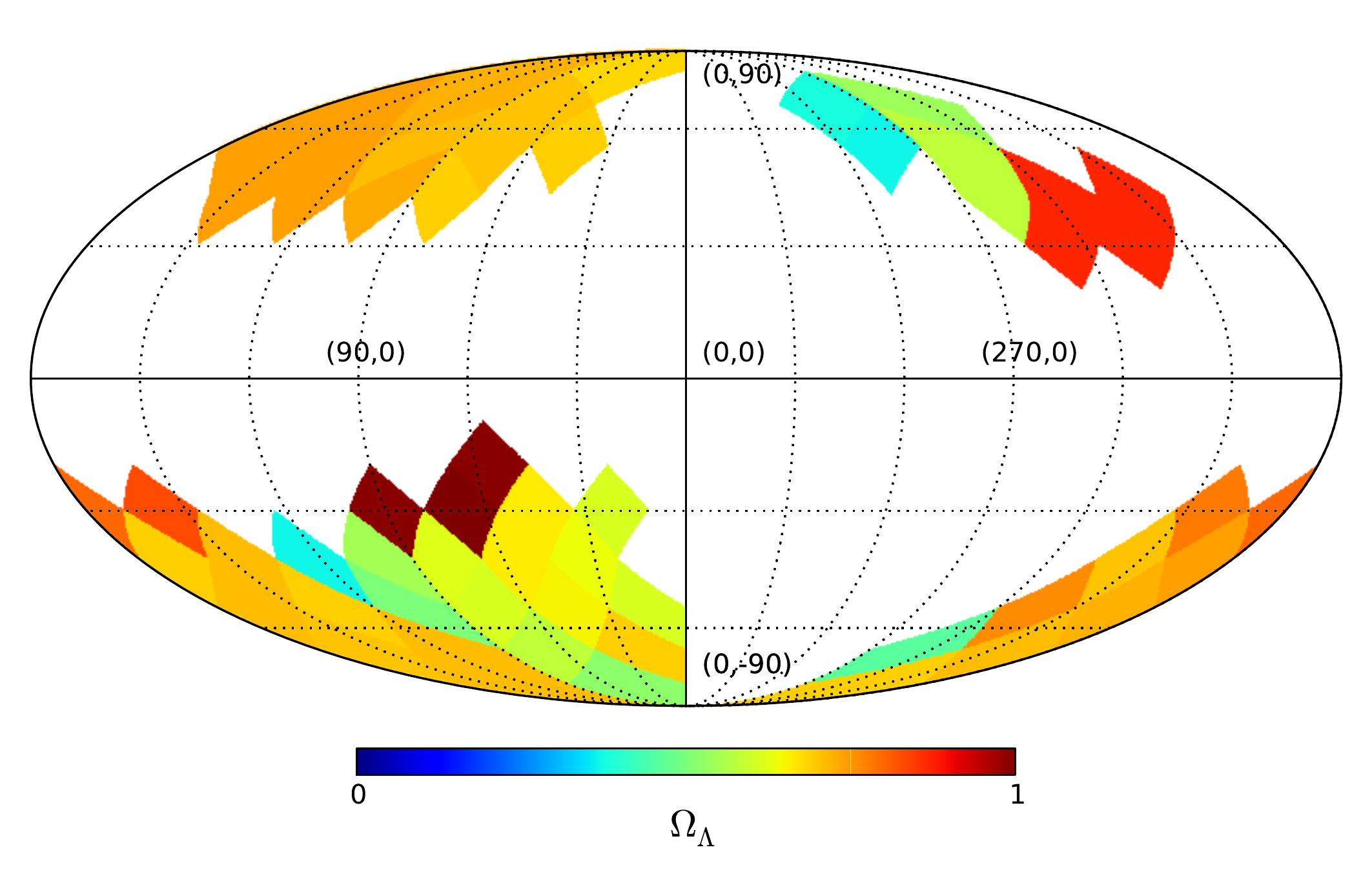}
\includegraphics[scale=0.4]{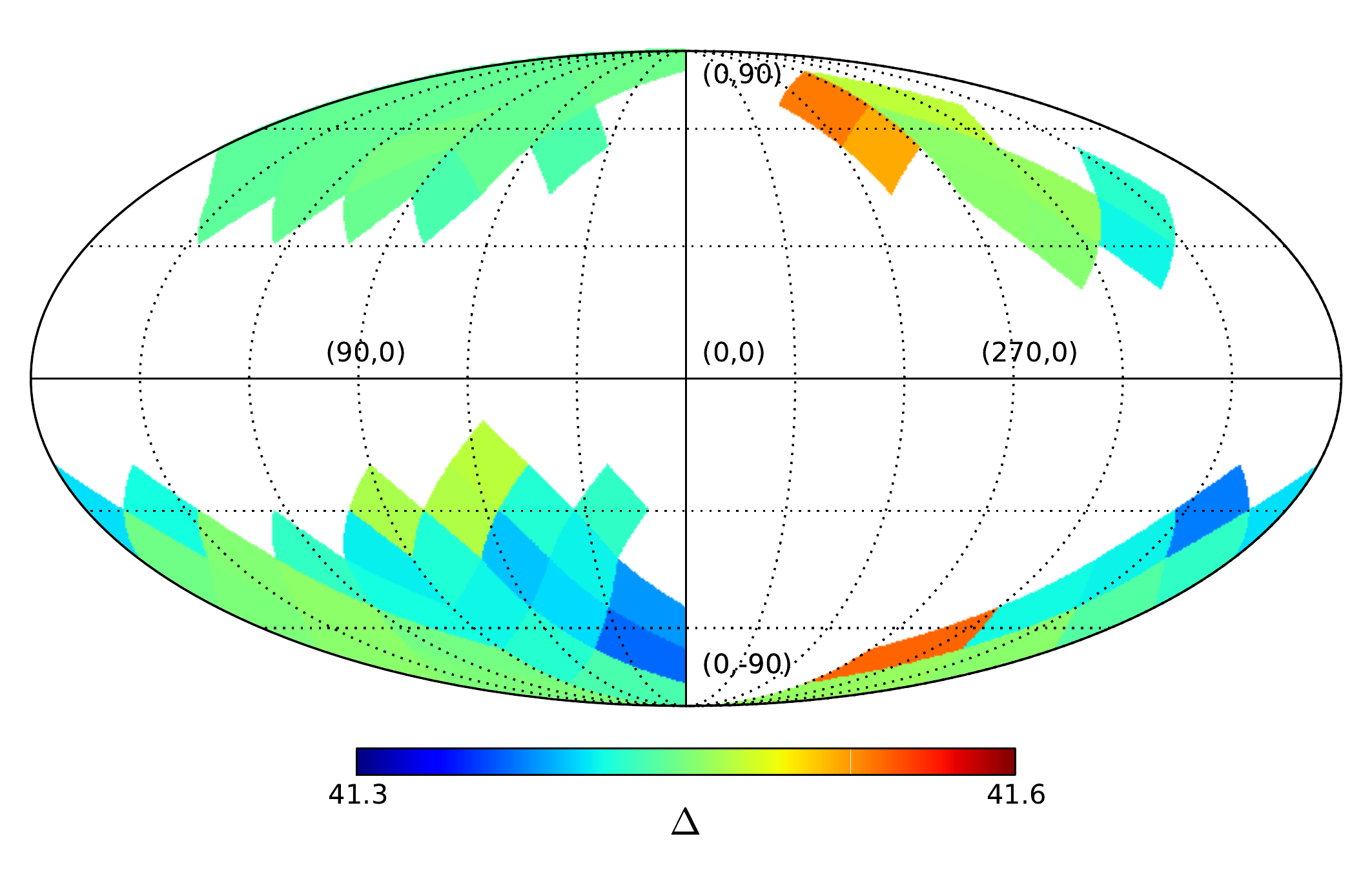}
  \caption{Best-fit values for $\Omega_{\Lambda}$ (left) and the correction parameter $\Delta$ (right) in different directions on the sky. Each pixel shows results that have been determined considering the magnitude-redshift relation of all Union-2.1 supernovae with redshift $z\geq 0.2$ that
lie within an angle $\theta$ from the pixel center.  The cone opening angle $\theta$ assumes
the values $\frac{\pi}{2}$ (top), $\frac{\pi}{3}$ (middle) and $\frac{\pi}{6}$ (bottom). 
The white regions indicate the directions for which the sampling cone contains less than 25 SNe Ia. }
  \label{fig:omega}
\end{figure*}

Sky maps of the best-fit values for $\Omega_\Lambda$ (left) and $\Delta$ (right) are shown in
Figure \ref{fig:omega} for three different cone opening angles (from top to bottom: $\theta=\frac{\pi}{2}$, $\frac{\pi}{3}$ and $\frac{\pi}{6}$ radians). These have been obtained using all the Union2.1 SNe with redshift $z\geq 0.2$.
White pixels indicate the directions (mostly located around the Galactic equator) in which the corresponding cone contains less than $25$ SNe Ia.
These directions are excluded from all statistical analyses because they are associated with extremely large errors in the fitted parameters. Of course their number increases  
with decreasing the opening angle of the sampling cone.  
Similarly, the size of fluctuations in the best-fit values for $\Omega_\Lambda$ and $\Delta$ increases
with reducing $\theta$. 

\subsubsection{Most discrepant directions}
Although Figure \ref{fig:omega} gives a first visual impression of the local best-fit parameters, it does not take into account the non-uniform sky coverage of the Union2.1 data set. For a given opening angle, different directions on the celestial sphere are generally associated with very different numbers of SNe Ia. This strongly influences the uncertainty of the best-fit values. 

In order to single out the most discrepant directions in a statistically meaningful way, we assume the null hypothesis that the Universe follows the cosmological principle and there are no angle-dependent systematic effects plaguing the Union2.1 sample. 
For each pixel we then evaluate the $\chi^2$ target function fixing the 
free parameters at the values $\hat\Omega_\Lambda=0.70$ and $\hat \Delta=41.428$ that provide the best-fit solution for the complete Union2.1 sample. However, only the SNe within the sampling cone are used to calculate the $\chi^2$ value that we denote by $\hat\chi^2$.
Finally, we estimate the probability $P$ that random noise could generate a
$\chi^2$ value exceeding $\hat\chi^2$.
Assuming Gaussian errors, this probability coincides with the cumulative chi-square distribution
function evaluated at $\hat\chi^2$:
\begin{equation}
\label{eq:pchi}
P=\frac{1}{2^{\nu/2}\Gamma(\nu/2)}\int_{\hat{\chi}^2}^{\infty} t^{\nu/2-1}e^{-t/2}dt\;,
\end{equation}
where $\nu$ is the number of degrees of freedom -- i.e. the number of SNe Ia used in the fitting procedure minus two (the number of free parameters). 
We adopt the $P$ value as a measure of how well the all-SNe best-fit parameters also describe the SN data in a specific direction on the sky. Consequently we identify the most discrepant direction 
(i.e.  the direction showing the most statistically significant deviation from isotropy) 
with the pixel showing the smallest $P$ value.
It is worth stressing that this is not necessarily the direction in which the Universe (or the SN data) might present the strongest intrinsic anisotropy but only the direction in which, given the current data, we can measure the most meaningful deviation in terms of the signal-to-noise ratio.

\begin{figure}
\begin{center}

\includegraphics[scale=0.4]{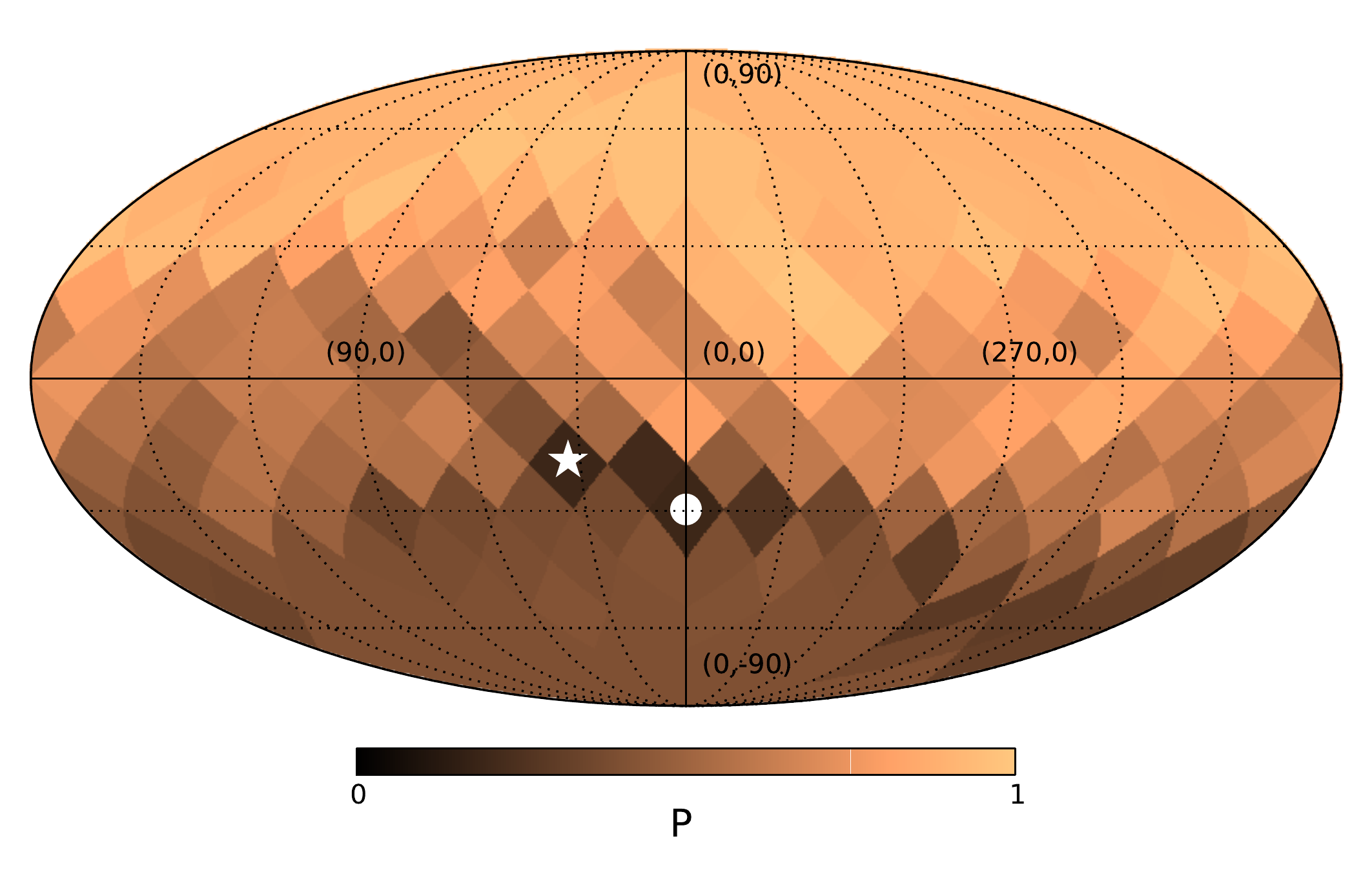}
\includegraphics[scale=0.4]{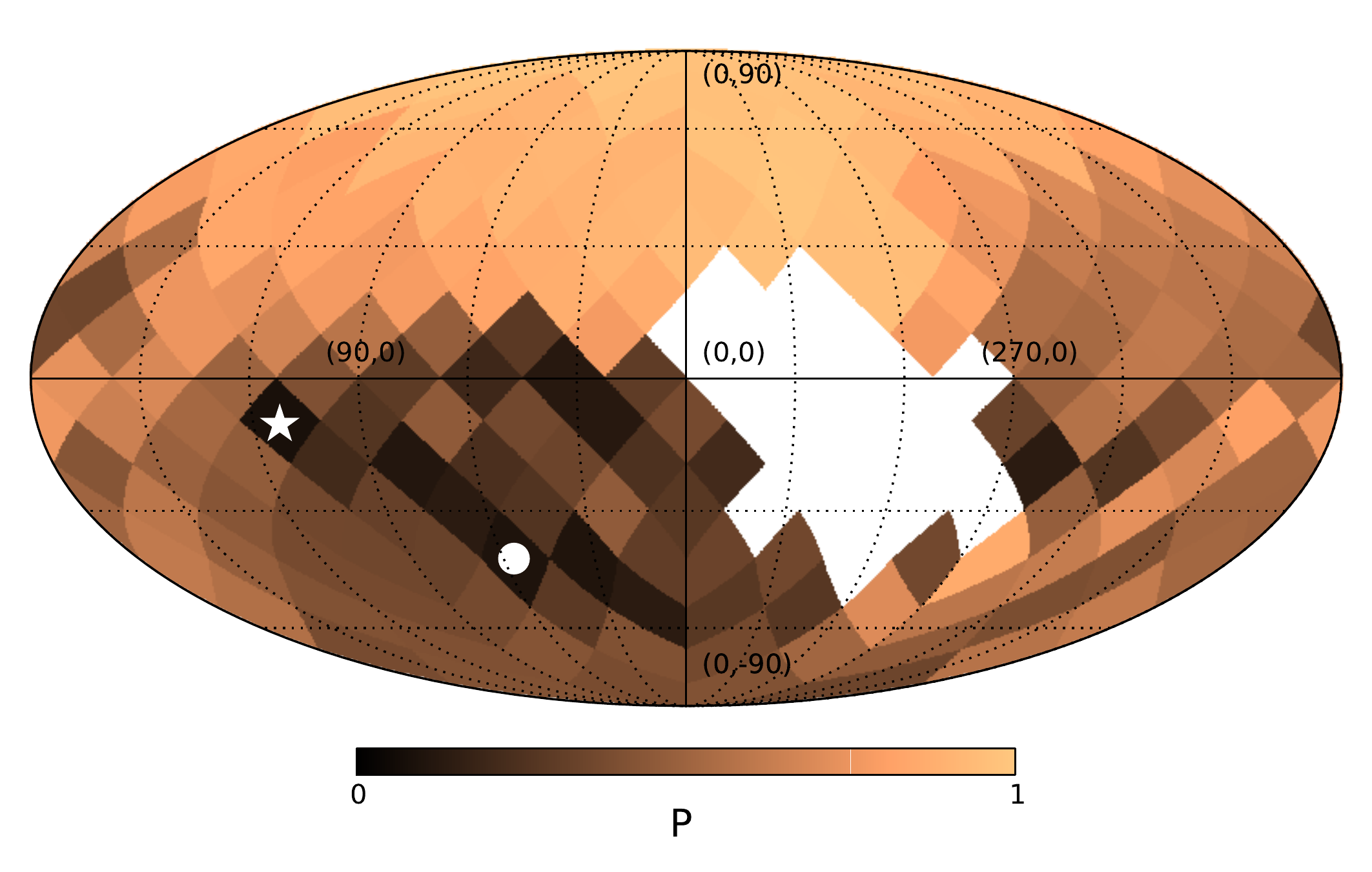}
\includegraphics[scale=0.4]{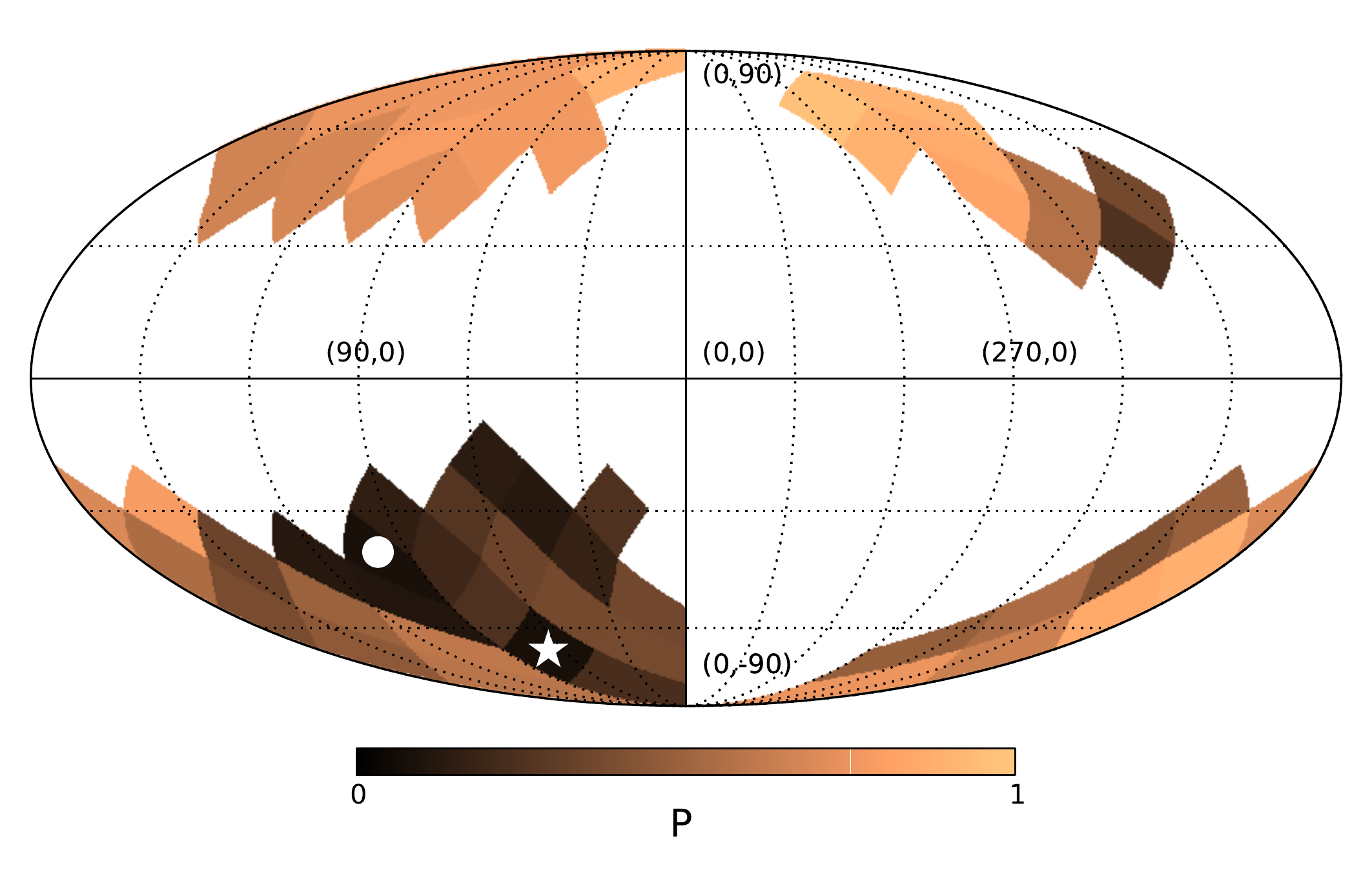}
  \caption{Maps of the $P$ value (the estimated likelihood of getting larger deviations than in the data
  due to random fluctuations under the null hypothesis that
  the SN Ia Hubble diagram is isotropic) for different cone opening angles (from top to bottom, $\theta=\frac{\pi}{2}$, $\frac{\pi}{3}$ and $\frac{\pi}{6}$). The direction with the smallest value of $P$ in each map is marked with a star and the second-most discrepant direction is highlighted with a circle.
  The white regions denote the pixels for which the cone contains less than 25 SNe Ia and are excluded from the statistical analysis.}
  \label{fig:pmap}
    
 \end{center}
\end{figure}

Maps of the $P$ value are plotted in Fig. \ref{fig:pmap} for the three different cone opening angles. The most (second-most) discrepant directions are highlighted with a star (circle) in each panel. Further information
is provided in Table \ref{t:prob} which gives the $P$ value, the coordinates and the number of SNe Ia 
associated with the most and the second-most discrepant directions together with the local best-fit parameters. The motivation for showing two directions per map is as follows: i) the difference between their $P$-values is small (see Table \ref{t:prob}), ii) the covariance matrix provided by the SCP is likely to be a noisy estimate, and iii) neglecting off-diagonal covariances switches the order between these directions for $\theta=\frac{\pi}{3}$.

\begin{figure*}
\includegraphics[scale=0.3]{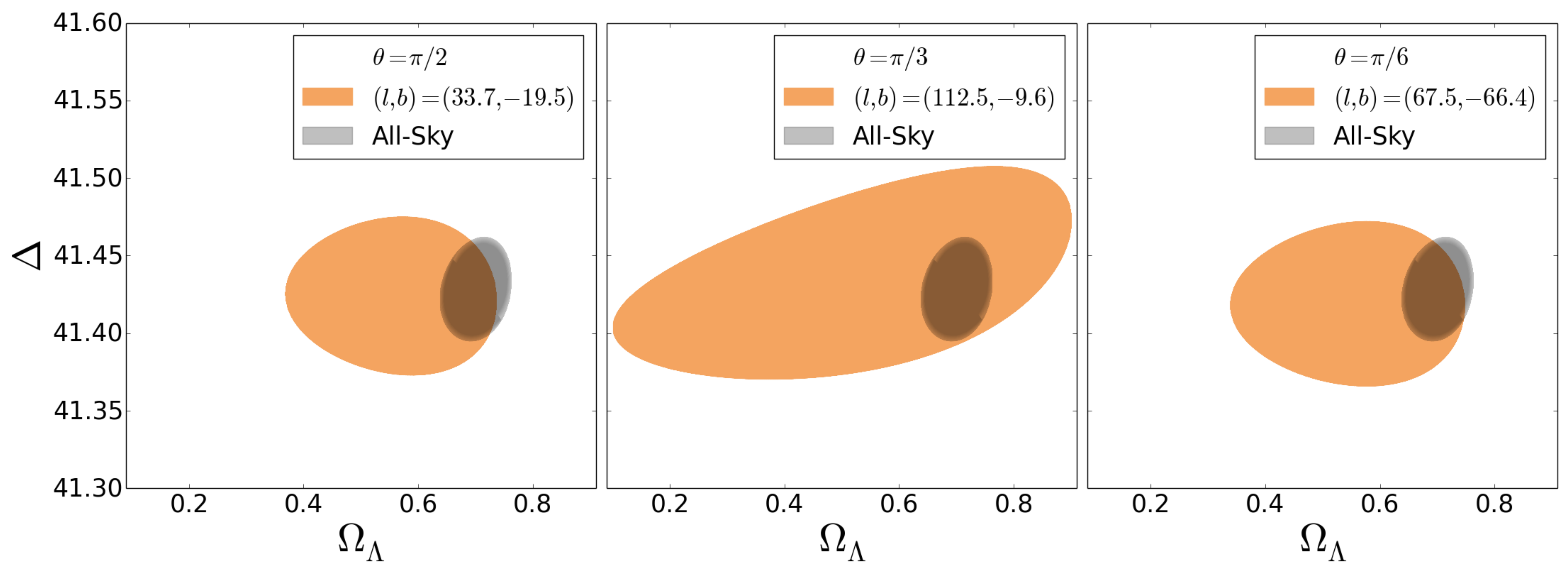}
  \caption{Formal $1\sigma$ ($\Delta \chi^2 \leq 2.30$) confidence regions from the all-SNe fit (gray) and the localized fits along the most discrepant directions (orange). From left to right the cone opening angle assumes the values $\theta=\frac{\pi}{2}$, $\frac{\pi}{3}$ and $\frac{\pi}{6}$.}
  \label{fig:conf_int_dis}
\end{figure*}

Intriguingly, the most discrepant directions obtained with the three cone opening angles lie close to each other. 
Also the best-fit parameters are quite similar
(let us not forget, however, that the maps with different $\theta$ are not independent as they use the same SNe and that there is significant overlap between the most discrepant cones). 
In Figure \ref{fig:conf_int_dis} we compare the formal\footnote{I.e. obtained assuming independent Gaussian errors. The limitation of this approach is discussed in detail in Section \ref{sec:mc}.  } $1\sigma$ confidence regions ($\Delta \chi^2 \leq 2.30$) obtained from the all-SNe fit against those derived from the local fits along the most discrepant directions. In all cases, the tension between the local and the global fits is marginal and the formal $1\sigma$ regions always overlap.

Visual inspection of Figure \ref{fig:pmap} shows a striking contrast between the $P$ values
measured in the Northern and the Southern Galactic Hemispheres (hereafter NGH and SGH, respectively), although there is no tension between the luminosity-distance relation in the two hemispheres (see Table \ref{t:hemis}). The discrepancy in the $P$ values is mainly due to the fact that the Union2.1 uncertainties in the distance-moduli are on average 30 per cent larger in the NGH. Consequently, the reduced chi-square $\hat{\chi}^2/\nu$ tends to be smaller in the NGH although there are many more SNe in the SGH (227 vs 123) to drive the fit results for SNe with $z\geq0.2$ closer to the SGH results.

%\clearpage
\setlength{\tabcolsep}{3pt}
\begin{table}
\begin{center}
\caption{Most discrepant directions\label{t:prob}}

\begin{tabular}{ccccccc}
\tableline\tableline
$\theta$ & $(l,b)$& $\Omega_\Lambda$& $\Delta$ &$\hat{\chi}^2/\nu$ &$P$ & $N$ \\
\tableline
$\bigstar$ $ \pi/2$ & (33.7,-19.5)& $0.58_{-0.13}^{+0.11}$ & $41.424^{+0.077}_{-0.072}$ & 1.09 & 0.192 & 128 \\
$\bullet$ $ \pi/2$ & (0.0,-30.0) & $0.61_{-0.10}^{+0.09}$ & $41.441^{+0.066}_{-0.067}$ & 1.08 & 0.197 & 161 \\
\tableline
$\bigstar$ $ \pi/3$ & (112.5,-9.6)& $0.60_{-0.31}^{+0.22}$ & $41.439^{+0.130}_{-0.110}$ & 1.20 & 0.086 & 74 \\
$\bullet$ $ \pi/3$ & (56.2,-41.8)& $0.59_{-0.15}^{+0.11}$ & $41.424^{+0.081}_{-0.072}$ & 1.15 & 0.101 & 118 \\
\tableline        
$\bigstar$ $ \pi/6$ & (67.5,-66.4)& $0.58_{-0.15}^{+0.12}$ & $41.419^{+0.084}_{-0.074}$ & 1.18 & 0.081 & 100 \\
$\bullet$ $ \pi/6$ & (101.2,-41.8)&$0.55_{-0.29}^{+0.21}$ & $41.408^{+0.124}_{-0.106}$ & 1.20 & 0.085 & 73 \\
\tableline
\end{tabular}
\tablecomments{Galactic coordinates $(l,b)$ and $P$ values characterizing the most (stars) and the second-most (circles) discrepant directions for different cone opening angles, $\theta$. Also reported are the number of SNe Ia in the cones, $N$, the best-fit values for $\Omega_{\Lambda}$ and $\Delta$ (in mag) and the ratio $\hat{\chi}^2/\nu$.}
\end{center}
\end{table}

%\clearpage
\begin{table}
\begin{center}
\caption{Results of NGH and SGH fitting \label{t:hemis}}

\begin{tabular}{lccccc}
\tableline\tableline
 &$\Omega_{\Lambda}$ & $\Delta$ &$\chi^{2}/\nu$ &N&$\bar{\sigma}_{\mu}$  \\
\tableline
NGH&$0.70^{+0.07}_{-0.09}$ & $41.443^{+0.088}_{-0.084}$ & 0.82&123&0.30   \\

SGH &$0.68^{+0.06}_{-0.08}$ & $41.441^{+0.060}_{-0.056}$ &1.01 &227&0.23\\
\tableline
\end{tabular}
\tablecomments{Best-fit values 
obtained using SNe in the NGH and SGH, separately. Also reported are the corresponding reduced chi-square, $\chi^{2}/\nu$, the number of SNe Ia with $z\geq0.2$ considered for the fit, $N$, and their average distance-modulus uncertainty, $\bar{\sigma}_{\mu}$.}
\end{center}
\end{table}

\subsubsection{Monte Carlo analysis}\label{sec:mc}
Taken at face value, the probabilities $P$ associated with most discrepant directions (see Table \ref{t:prob}) are moderately significant. However, assuming Gaussian errors is a strong limiting factor. Also, the size of the errorbars in the distance modulus (and the off-diagonal covariances) provided in the Union2.1 catalog might be inaccurate and, as a consequence, inference based on the $\chi^2$ statistic might be biased. 
For these reasons, we re-evaluate the statistical significance of the anisotropies using a more robust Monte Carlo method. 

In order to assess the impact of random errors and account for the non-uniform angular
distribution of the Union2.1 sample, we build 1000 mock catalogs by randomly shuffling the distance moduli of the Union2.1 SNe. 
In practice, we assign the distance modulus, its uncertainty and the redshift of a SN Ia to the angular 
position of another (random) SN Ia.
Each mock catalog thus contains exactly the same number of SNe as the original Union2.1 sample and has exactly the same SN sky distribution. Moreover, all possible anisotropies should be erased by the shuffling procedure while the statistical properties of the distance moduli and their uncertainties are unchanged with respect to the observational data. Therefore our mock catalogs form an ensemble of realizations mimicking an isotropic Universe but having the same statistical properties as the actual Union2.1 data.

We treat the mock catalogs as the real data and identify the two most discrepant directions in each of them using the $P$-value-based method for the three different cone opening angles. We then compute the fraction, $f_0$, of the realizations in which the most (or the second most) discrepant direction is associated with a $P$
value which is smaller than the observed ones reported in Table \ref{t:prob}. 
For the most (second-most) discrepant directions we find that $f_0=0.569, 0.623$ and $0.674$ (0.473, 0.550, 0.490) for $\theta=\frac{\pi}{2}$, $\frac{\pi}{3}$ and $\frac{\pi}{6}$, respectively. Purely based on this signal-to-noise criterion,
we conclude that no statistically significant anisotropy can be detected in the Union2.1 sample. 

\subsubsection{Alignment with the CMB dipole}
Although the Monte Carlo test shows that random chance in an isotropic universe can easily produce most discrepant directions with lower $P$ values than we found analyzing the actual data, the  observed anisotropies present a characteristic feature which is worth being discussed.

\begin{figure}
\includegraphics[scale=0.27]{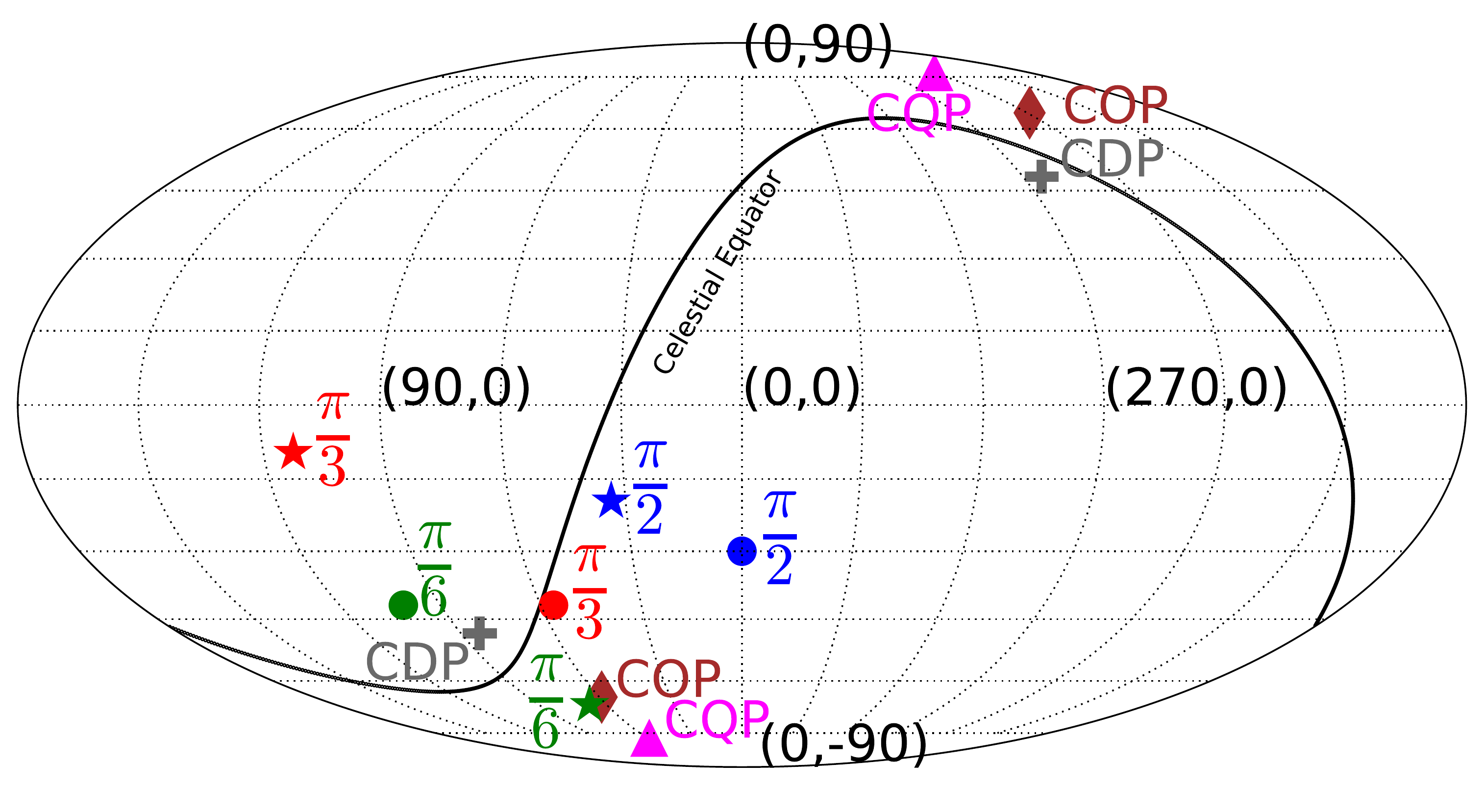}
  \caption{The most (star) and the second-most (circle) discrepant directions in the magnitude-redshift relation of SNe Ia (for three cone opening angles $\theta=\frac{\pi}{2}$, $\frac{\pi}{3}$ and $\frac{\pi}{6}$) obtained in this work are compared with the directions of the CMB dipole (CDP), quadrupole (CQP) and octopole (COP) from \citet{planck}. The black solid curve denotes the celestial equator.}
  \label{fig:aniso}
\end{figure}

The temperature distribution in the CMB presents a strong dipole anisotropy which is usually interpreted
as due to our motion with respect to the CMB rest frame towards the direction with Galactic
coordinates $(263^\circ\!\!.99\pm0.14,48^\circ\!\!.26\pm0.03)$ \citep{planck_dop}.
Figure \ref{fig:aniso} shows that the most discrepant directions we obtained from the Union2.1 sample closely align with the axis of the CMB dipole (CDP) in the SGH opposite to our motion with respect to the CMB rest frame (hereafter CDP-South). Assuming that the redshifts of the SNe Ia in the Union2.1 compilation have been correctly transformed to the CMB rest frame, there is no obvious reason for explaining the origin of this alignment.
As already mentioned in the Introduction, the CMB quadrupole (CQP) and octopole (COP) are also
closely aligned with the CDP \citep[][see Figure \ref{fig:aniso}]{planck, schwarz04,copi10,copi13}. It is yet unclear whether these alignments are a statistical fluke or a signature of new physics. Anyway, our study shows that the magnitude-redshift relation of SNe Ia with $z\geq0.2$ tends to be different in the same direction (albeit the difference is detected with low signal-to-noise ratio).
Other authors have reported similar results using the Union compilations \citep{cooke,antoniou,li}. 

The debate on the physical relevance of the CMB anomalies opened up a discussion in the literature about the legitimacy and validity of ``a posteriori" analyses in which tailored statistical tests are designed and hand picked after noticing the peculiarities in the data. A widespread point of view states that in a large dataset it is always possible to isolate some ``strange" features \citep[e.g.][]{bennett11}.
To minimize the pitfalls of a posteriori reasoning, we focus on the well established CDP and do not
consider the CQP and COP any further.

\begin{figure*}
\includegraphics[scale=0.32]{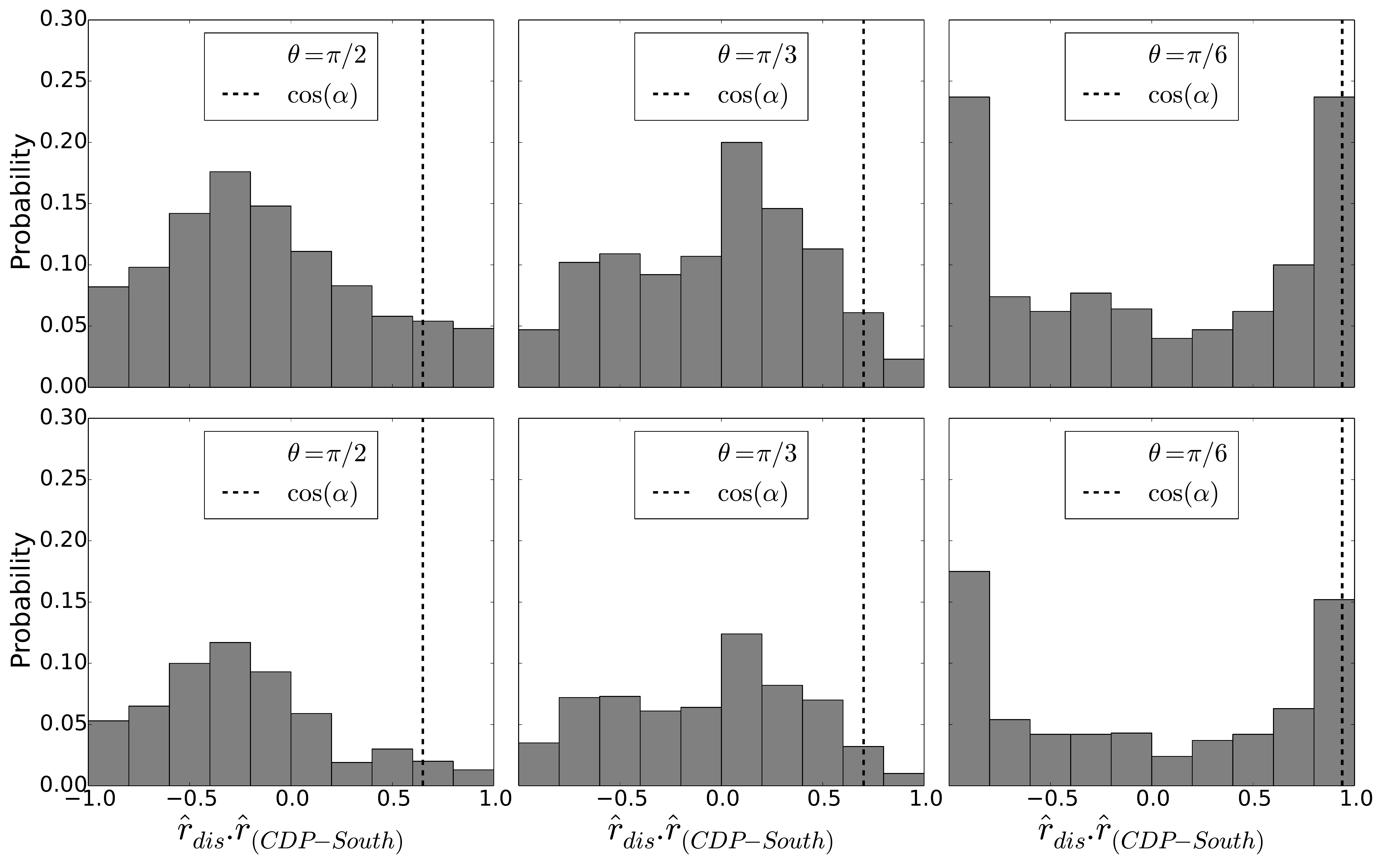}
  \caption{Top: Probability distribution of the cosine of the angle between the direction of the CMB dipole in the SGH, $\hat{r}_{CDP-South}$, and the most discrepant direction of the Hubble diagram of SNe Ia, $\hat{r}_{dis}$, determined using the Monte Carlo realisations introduced in Section \ref{sec:mc}. Bottom: As above but with the additional condition that the most-discrepant direction is associated with a smaller $P$ value than for the Union2.1 measurement. From left to right, the panels refer to the cone-opening angles $\theta=\frac{\pi}{2}$, $\frac{\pi}{3}$ and $\frac{\pi}{6}$. The cosine of the observed angular separation in the Union2.1 sample, $\alpha$, (reported in Table \ref{t:probs}) is shown as a dashed line.}
  \label{fig:prob_hist}
\end{figure*}

We thus proceed to quantify the probability that the most discrepant directions (defined in terms
of the $P$ value as above) form a given angle with the CDP under the null hypothesis of 
an isotropic magnitude-redshift relation.
In order to account for the non-uniform sky distribution of the Union2.1 sample (especially for the
SDSS-II stripe which is close to the CDP-South) we use the Monte Carlo realizations introduced
in Section \ref{sec:mc}.
Figure \ref{fig:prob_hist} shows the resulting probability distribution for the cosine of the angle
between the most discrepant direction and the axis of the CDP-South. Our measurements from the Union2.1 data are indicated by vertical dashed lines. 
The second column in Table \ref{t:probs} reports the fraction of Monte Carlo realisations, $f_{1}$, showing a better alignment than our measurement. Our results suggest that it is rather unlikely to get an alignment as strong as the observed one under the null hypothesis of an isotropic magnitude-redshift relation. In fact, considering the most discrepant direction for $\theta=\frac{\pi}{2}$, only 8.5 per cent of the Monte Carlo realisations show a smaller separation angle than observed and this reduces to 4.5 percent for $\theta=\frac{\pi}{3}$. Note that for $\theta=\frac{\pi}{3}$ and $\frac{\pi}{6}$, the second-most discrepant directions are even better aligned with CDP-South. In these cases $f_1\simeq 0.01$.

The test above is blind to the statistical significance of the most discrepant directions. In order to account for this, we compute the fraction of Monte Carlo realisations, $f_{2}$, for which the most discrepant directions are at least as significant as the measured ones (in terms of the $P$ value) and are also better aligned with the CDP-South. The third column in Table \ref{t:probs} shows that for the most discrepant directions this probability is smaller than 4.5 percent for all the opening angles which means the null hypothesis of an isotropic magnitude-redshift relation
should be rejected at the 95 per cent confidence level. The value of $f_2$ reduces to a fraction of a percent when considering the second-most discrepant direction for $\theta=\frac{\pi}{3}$ and $\frac{\pi}{6}$.

The measured anisotropy could be due to a statistical fluke, to systematics in the SNe data (or error bars), to the presence of localized large scale structures, or even a sign of the failure of the cosmological principle. To further investigate its properties,
we repeat the analysis along the most discrepant directions after slicing the SNe data in five redshift
bins ($0.2\leq z <0.3$, $0.3\leq z <0.4$,  $0.4\leq z <0.6$,  $0.6\leq z <0.9$ and $z\geq0.9$). Regrettably, due to the low number of SNe in each bin, the formal $1\sigma$ errors span most, if not all, the parameter space $0\leq\Omega_{\Lambda}\leq 1$. Therefore no meaningful statements can be made regarding the
variations of the best-fit cosmological parameters along the most-discrepant
directions. In terms of signal-to-noise ratio, however, the redshift range
$0.4\leq z <0.6$ clearly emerges as the most discrepant one for all the cone opening angles 
($P=0.033, 0.021$ and $0.016$ for $\theta=\pi/2,\pi/3$ and $\pi/6$, respectively).

Given the current sparsity of the data, no firm conclusion can be drawn except from the fact that
there seems to be a moderately statistically significant (2-3$\sigma$) anisotropy in the magnitude-redshift
relation of SNe Ia close to the direction
opposite to our motion with respect to the CMB rest frame. It is worth remembering that, in the Union2.1 compilation, most of the SNe Ia surrounding the CDP-South 
come from the SDSS-II stripe. 
Further investigations are thus needed to clarify the relation between the 
CMB dipole axis, our motion, and the way SNe data around this direction are treated. 

\begin{table}
\begin{center}
\caption{Alignment with the CMB dipole \label{t:probs}}

\begin{tabular}{cccc}
\tableline\tableline
$\theta$ & $\alpha$ & $f_{1}$& $f_{2}$\\
\tableline
$\bigstar$ $ \pi/2$ &$49^{\circ}\!\!.4$ &0.085&0.027\\
$\bullet$ $ \pi/2$ &$64^{\circ}\!\!.3$ &0.152&0.052\\
\tableline
$\bigstar$ $ \pi/3$ &$45^{\circ}\!\!.5$ &0.045&0.021\\
$\bullet$ $ \pi/3$ &$20^{\circ}\!\!.5$ &0.008&0.002\\
\tableline
$\bigstar$ $ \pi/6$ &$20^{\circ}\!\!.1$ &0.064&0.045\\
$\bullet$ $ \pi/6$ &$13^{\circ}\!\!.7$ &0.010&0.006\\

\tableline
\end{tabular}
\tablecomments{Angular separation, $\alpha$, between the direction of the CMB dipole in the SGH 
(CDP-South) and the most (stars) and the second-most (circles) discrepant directions for the maps based on the Union2.1 data with different cone opening angles, $\theta$. 
The probability of measuring a value smaller than $\alpha$ in random realisations of a isotropic
magnitude-redshift relation is indicated with $f_1$ while $f_2$ also accounts for the condition that
the most (second-most) discrepant direction is associated with a smaller $P$ value than for the Union2.1 measurement.
Both probabilities have been estimated with a Monte Carlo method (see the main text for the details).}
\end{center}
\end{table}

\section{Conclusions and future perspectives}
We presented a simple but powerful method for investigating the isotropy of cosmic acceleration traced by Type Ia SNe with different angular resolution, $\theta$.
The key idea is to consider all the SNe contained within a cone with vertex located at the origin
of the Galactic coordinate system and with apex angle $2\theta$.
``Local cosmological parameters" are derived by fitting the magnitude-redshift relation of the SNe in the cone with a theoretical relation. The cone direction is then changed so that to cover the entire sky. 
Our cone-analysis method takes into account the mean variation of the SNe Ia correction parameters over different directions, and yields all-sky maps of the best-fit cosmological parameters. 

Although a large data set with a uniform sky distribution is required for a thorough investigation of isotropy, we provided an example of the potential of our method by applying it to the SNe Ia with redshift $z\geq0.2$ in the Union2.1 compilation.
Assuming a flat Universe in the context of the standard cosmological model, we fitted the magnitude-redshift relation by varying
the density parameter of the cosmological constant, $\Omega_{\Lambda}$, and a parameter , $\Delta$, including the effect of both the Hubble constant and the mean SNe Ia correction parameters. We used a HEALPix grid to discretise the celestial sphere and obtained sky maps for $\Omega_{\Lambda}$ and $\Delta$ considering three different cone-opening angles $\theta=\frac{\pi}{2}$, $\frac{\pi}{3}$ and $\frac{\pi}{6}$. 

We ranked the pixels in each map in terms of a $P$ value derived from the $\chi^2$ distribution
and which measures how much the local fits differ from the cosmology determined using the entire
Union2.1 sample (in a signal-to-noise sense).
We thus found the most discrepant directions (two per cone opening angle).  
Finally, we used a Monte Carlo method to estimate the statistical significance at which we could reject
the null hypothesis that the magnitude-redshift relation of SNe Ia is isotropic based on the properties
of the most discrepant directions.
We found that random fluctuations can easily produce deviations from isotropy with smaller $P$ values
than measured in the Union2.1 data. Therefore, the null hypothesis cannot be rejected at any 
meaningful confidence level based on signal-to-noise arguments alone.
However, if we also consider that the detected anisotropies in the Union2.1 sample align well with CMB dipole axis in the
Southern Galactic Hemisphere, we find that the null hypothesis should be rejected at the 97.3, 97.9 and 95.5 per cent confidence level for opening angles $\theta=\frac{\pi}{2}$, $\frac{\pi}{3}$ and $\frac{\pi}{6}$, respectively.

We conclude that, although the deviation from isotropy that we found is not very significant per se in terms of signal-to-noise ratio, its vicinity to the axis of the CMB dipole (which enters the pipeline to determine the SN redshift in the CMB rest frame) with 2-3$\sigma$ statistical significance requires further investigation both on the observational and on the theoretical sides. Note that other observations detected anisotropies in the same area of the sky. The statistical significance of the quadrupole--octopole alignment in the CMB is approximately 99 per cent \citep{planck}. On combination of the likelihoods between the CMB and SN Ia, the null hypothesis of isotropy should be rejected at the 99.98 per cent confidence level (approximately 3.5 Gaussian $\sigma$). In this paper we followed a conservative approach by only considering the SN Ia data.

\begin{figure}
\includegraphics[scale=0.27]{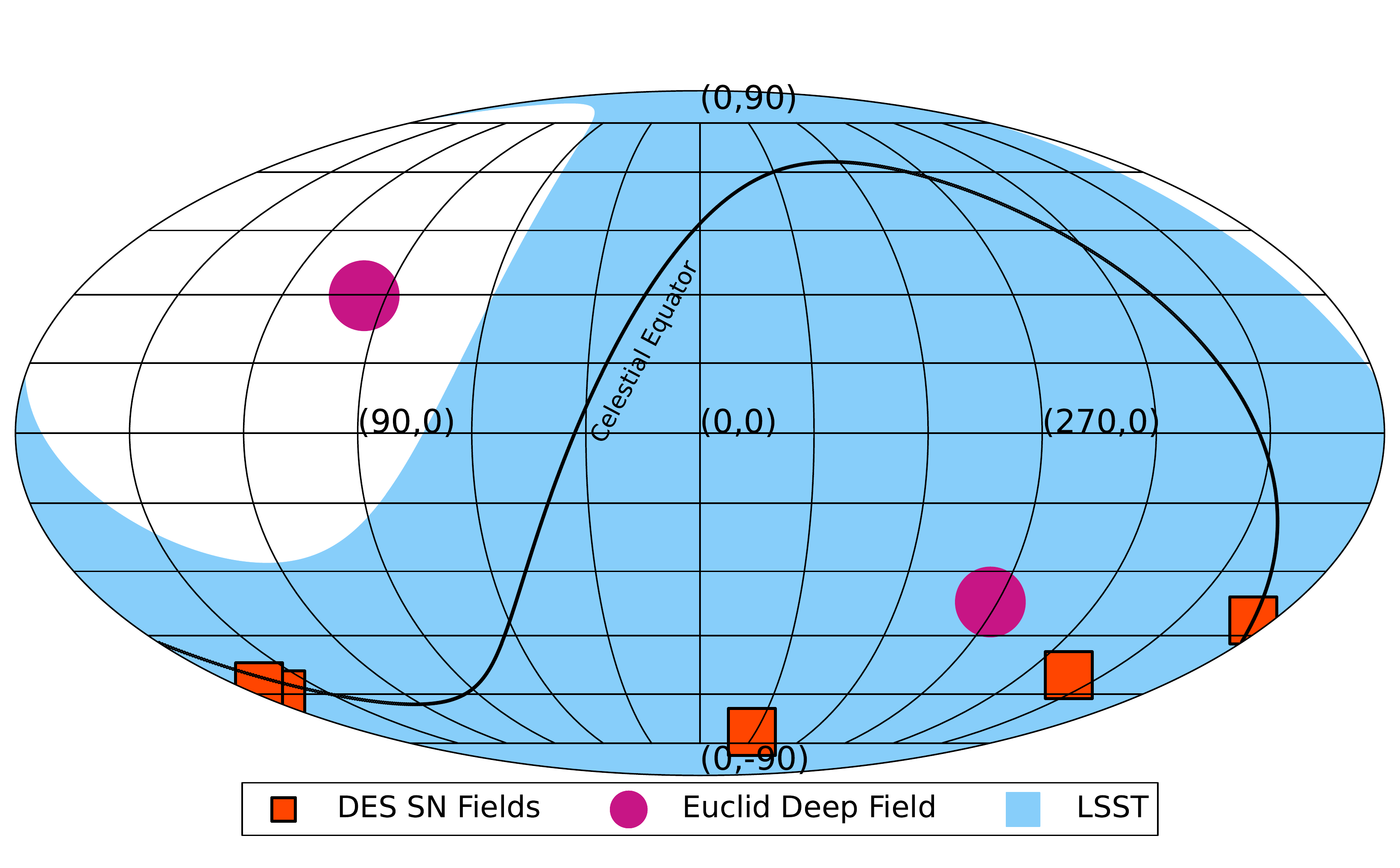}
  \caption{Survey footprints for the Euclid deep field, LSST main survey and the (likely) SNe fields of DES in the Galactic coordinate system. The size of the fields for DES and Euclid have been artificially magnified to ease readability. The black solid curve denotes the celestial equator.}
  \label{fig:surveys}
\end{figure}

This study should be repeated when larger data sets with more uniform sky coverage will be available. 
Several major current and future facilities have dedicated plans for studying the accelerated expansion of the universe using SNe Ia. For instance,
the Dark Energy Survey (DES) integrates a dedicated program that should detect around 4000 SNe Ia in the redshift range $0.05<z<1.2$ \citep{des}. 
Similarly, 
the Euclid mission includes a SNe survey within two deep fields each covering around 20 deg$^2$ and is expected to discover about 3000 SNe Ia out to $z \approx 1.2$ \citep{euclid}. 
However, both these surveys will only provide SN data in relatively small regions of the sky 
(see Figure \ref{fig:surveys}) and
the most promising perspective for isotropy tests of the Hubble diagram comes from 
 the Large Synoptic Survey Telescope (LSST). While its use for a SN-dedicated survey on a limited area of sky will be able to deliver as many as $140,000$ SNe Ia (in 10 years) with very precisely measured light curves, in its normal operating mode (due to its rapid cadence), LSST will discover around $250,000$ SNe Ia per year in the redshift range $0.45<z<0.7$ and across a large fraction of the sky \citep{lsst}.  Finally, the Panoramic Survey Telescope \& Rapid Response System  \citep[Pan-STARRS,][]{pan} which is observing the Northern part of the sky will complement the above mentioned surveys.
In summary, exciting perspectives to test the isotropy of the magnitude-redshift relation of SNe Ia
with unprecedented accuracy will open up within the next two decades.

\section*{Acknowledgments}
We thank the referee for providing constructive comments on the manuscript. BJ thanks Ryan Cooke for help with finding some of the SNe Ia coordinates, Dominik Schwarz and Marek Kowalski for constructive comments and Douglas Applegate for very useful discussions. BJ was supported through a stipend from the International Max Planck Research School (IMPRS) for Astronomy and Astrophysics at the Universities of Bonn and Cologne. CP acknowledges support from the Deutsche Forschungsgemeinschaft through the Transregio 33 ``The Dark Universe''.

\bibliography{biblio}

\end{document}